\documentclass[prb,twocolumn,superscriptaddress,showpacs]{revtex4}

\usepackage{epsfig,amsmath,amssymb,latexsym}

\newcommand{\beq}{\begin{eqnarray}}
\newcommand{\eeq}{\end{eqnarray}}
\usepackage{subfigure}

\begin{document}
\newcommand{\Go}{\ensuremath{\textrm G_{0}}}
\newcommand{\brho}{\mbox{\boldmath$\rho$}}

\title{\textit{Ab initio} study of spin-dependent transport in carbon nanotubes with iron and vanadium adatoms}
\preprint{1}
\author{Joachim A. F\"{u}rst}
\email[Corresponding author: ]{jof@mic.dtu.dk} 
\affiliation{MIC -- Department of Micro and
Nanotechnology, NanoDTU, Technical University of Denmark, DK-2800 Kongens Lyngby, Denmark}
\affiliation{Atomistix A/S, c/o Niels Bohr Institute, 2100 Copenhagen, Denmark}
\author{Mads Brandbyge}
\affiliation{MIC -- Department of Micro and Nanotechnology, NanoDTU, Technical University
of Denmark, DK-2800 Kongens Lyngby, Denmark}
\author{Antti-Pekka Jauho}
\affiliation{MIC -- Department of Micro and Nanotechnology, NanoDTU, Technical University
of Denmark, DK-2800 Kongens Lyngby, Denmark}
\affiliation{Laboratory of Physics, Helsinki University of Technology, P. O. Box 1100, FI-02015 HUT, Finland}
\author{Kurt Stokbro}
\affiliation{Department of Computer Science, Universitetsparken 1, DK-2100 Copenhagen \O, Denmark}
\date{\today}
\begin{abstract}
We present an \textit{ab initio} study of spin dependent transport in armchair carbon nanotubes with transition metal adsorbates, iron or vanadium. We neglect the effect of tube curvature and model the nanotube by graphene with periodic boundary conditions. A density functional theory based nonequilibrium Green's function method is used to compute the electronic structure and zero-bias conductance. The presence of the adsorbate causes a strong scattering of electrons of one spin type only. The scattering is shown to be due to coupling of the two armchair band states to the metal $3d$ orbitals with matching symmetry causing Fano resonances appearing as dips in the transmission function. The spin type (majority/minority) being scattered depends on the adsorbate and is explained in terms of $d$-state filling. The results are qualitatively reproduced using a simple tight-binding model, which is then used to investigate the dependence of the transmission on the nanotube width. We find a decrease in the width of the transmission dip as the tube-size increases.   
\end{abstract}

\pacs{
73.63.Fg,
73.63.-b
}
\maketitle
\section{Introduction}
Various carbon structures have attracted considerable attention in the last decade or so,
due to their potential use within future nanodevices \cite{Avouris2007}. Graphene, the single-atom thick
two-dimensional sheet of graphite is the basis for many carbon materials. Rolling up such a
sheet creates the one-dimensional single-walled carbon nanotube (SWNT) and wrapping up a
sheet makes 0D fullerenes. These materials have a wide range of remarkable electronic
properties. Recently graphene ribbons  -- cut graphene sheets -- were used to make the first
transistor carved in graphene \cite{transistor} which is stable at room temperature. 

Not only pushing the limits within electronics, these carbon materials may lay the ground for
the emerging field of spintronics \cite{prinz,wolf,Zutic}, where the functionality of the device is based on the spin degree of freedom. While graphene and SWNT's are not inherently magnetic, introducing
defects\cite{Lehtinen2004,kumazaki,Duplock2004}, impurities or boundaries\cite{okada,lee} may
alter this fact.  In recent years several studies have been carried out on SWNT/graphene -
transition metal adatom systems. First principles calculations have been employed to
determine equilibrium geometries and magnetic properties: Fagan $et$ $al.$ reported
electronic structure studies on single iron\cite{fagan3,fagan2,fagan1} and manganese
adatoms \cite {fagan2,fagan3} on zigzag SWNT's showing a total magnetic moment of the
systems and in most cases a magnetisation of the tube itself. A similar earlier study with additional
transition metal elements on graphene was published by Duffy \textit{et al.} \cite{Duffy1998}
reaching similar conclusions. Manganese dimers, trimers and wires on zigzag SWNT's\cite{fagan2} exhibit magnetic
moments close to free manganese. Yang $et$ $al.$ \cite{Yang} have proposed that iron and cobalt coated and filled SWNT's can
work as spintronic devices demonstrating a spin polarisation close to 90 \% at the Fermi
level as well as considerable magnetic moments. Interestingly, both semiconducting and
metallic SNWT's showed high spin polarisation. Kang $et$ $al.$ \cite{6} obtained similar
conclusions for various iron nanowire configurations inside armchair nanotubes \cite{6}. On the other hand, work by Kishi \textit{et al.} \cite{kishi} shows that wires of iron and cobalt lose their magnetic moments when adsorbed on armchair SNWT's. 
Filling of SWNT's with transition metals has been realised experimentally several years ago
\cite{monthioux}, which adds to the potential of these systems. A few studies have been published on electron transport in this context. Iron-SWNT's
junctions with C$_{60}$ molecules \cite{zhang} and pristine SWNT's \cite{guo} have been
proposed as magnetic tunnel junctions. First principles transport calculations yield
tunnel magnetoresistances of 11 \% and 40 \%, respectively, for these systems. A spin-polarised current has also been reported in theoretical works on nanoribbons with substitutional boron atoms
\cite{martins_dasilva}, where spin-dependent scattering is found.

Motivated by the reported spin-dependent scattering, we seek here to explain the physical mechanisms found in the numerical calculations. We investigate the influence on the transport properties of a single iron or vanadium atom adsorbed on a SWNT. By performing \textit{ab initio} spin-polarised transport calculations we demonstrate a spin-dependent scattering. The spin type being scattered is shown to depend on the type of adsorbate. By analysing the PDOS of the $3d$ orbitals of the adsorbates we find the scattering to be caused by coupling of the band states in the SWNT to these orbitals, resulting in Fano resonance phenomena. Using a simple tight-binding model we qualitatively reproduce these results, and also investigate the dependence of the transmission on tube size. 

The paper is organised as follows. Sec. \ref{sec_two} introduces our model system and the technical details of the calculations. The \textit{ab initio} results are presented in Sec. \ref{sec_three} and the Fano resonance is briefly introduced along with transmission eigenchannel analysis. In Sec. \ref{sec_four} we present a simple tight-binding model which is compared to the \textit{ab initio} results. Results for larger tubes are given based on the simple model.          
\section{System}\label{sec_two}
\begin{figure}
\begin{center}
\includegraphics[angle=0, width=1.0 \columnwidth,viewport=0 70 575 350,clip]{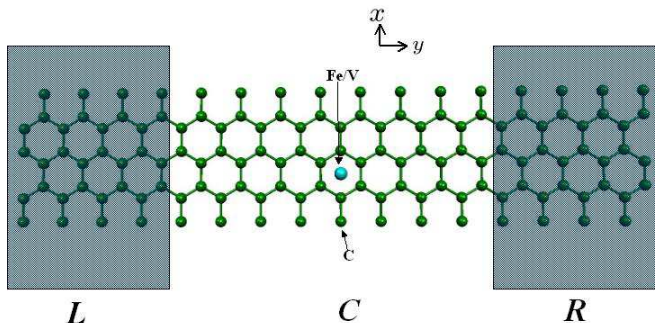}
\caption{(Colour online) The unit cell of the system. The sheet is cut in an armchair
structure along the \textit{y}-direction. The adatom, iron or vanadium is placed in the center.
The shaded areas marked \textit{L} and \textit{R} are the electrodes, and \textit{C} is the central region.
Periodic boundary conditions are imposed in the transverse direction (\textit{x}).}\label{ribbon}
\end{center}
\end{figure}

\begin{figure}
\begin{center}
\includegraphics[angle=-90, width=0.99 \columnwidth,viewport= 50 35 580 720,clip]{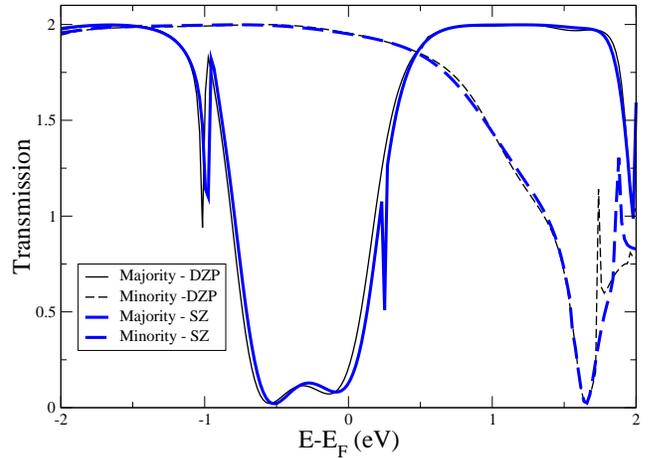}
\caption{(Colour online) The transmission of the vanadium adatom system using a single-zeta basis set for the whole system versus using double-zeta polarised for the adatom.}\label{basisset}
\end{center}
\end{figure}

The system used for transport calculations is shown in Fig. \ref{ribbon}. The graphene
sheet is cut in an armchair structure along the $y$-direction which results in a metallic
system. We employ periodic boundary conditions in the transverse $x$-direction. We have
used the {\it ab initio} pseudopotential density functional theory (DFT) as implemented in
the SIESTA code\cite{siesta} to obtain the electronic structure and relaxed atomic
positions from spin-polarised DFT. We employ the GGA PBE pseudopotential for exchange-correlation\cite{Perdew1996}. Our spin transport calculation is based
on the nonequilibrium Greens function method as implemented in the
TranSIESTA\cite{Brandbyge2002} code, extended to spin-polarised systems \cite{Datta1997}. However, we only
consider here the zero bias limit and focus on electron transmission close to the
Fermi energy. According to the TranSIESTA method \cite{Brandbyge2002} the system is
divided into left and right electrodes, marked \textit{L} and \textit{R} on Fig.
\ref{ribbon}, and a central region marked \textit{C}. A single iron or vanadium adatom is
placed in the middle of the central region. The electrodes both contain 32 atoms while the
central region consists of 64 sheet atoms. The adatom systems are relaxed using the CG
method with a force tolerance of 0.01~eV/\AA.  The carbon atoms were kept fixed during geometry optimisation. The relaxed position of the adatoms is in
the center of a hexagon at a distance of 1.73~{\AA} and 1.89~{\AA} from the sheet plane
for iron and vanadium, respectively. The period of the cell in the transverse $x$
direction is 8.52~{\AA}, which is then the smallest distance between adatoms in
neighbouring cells, and should be enough to prevent significant adatom -- adatom
interactions. We employ Gamma point sampling ($k=0$) in the periodic $x$ direction. The
system then corresponds to an (2,2) armchair nanotube described by the approximate
Graphene Sheet Model\cite{White2005}, that is, neglecting curvature effects. We will
return to the dependence of the electron transmission close to the Fermi energy on the
GSM-armchair tube width $(n,n)$ in Sec. \ref{sec_four}. 
The mesh cutoff value defining the real space grid was set
to 175 Ry. A single-zeta (SZ) basis set was used. The difference between a SZ basis
for all atoms and increasing the vanadium basis to double-zeta polarised (DZP)
can be seen on Fig. \ref{basisset}. The only qualitative difference between using DZP and SZ is the additional dip (a single point in the graph) at \textit{E} $\approx$ 0.25 eV in the SZ case. The reason for this not occuring for DZP is merely a matter of resolution. The dips will be addressed in the following two sections. In the remains of the paper we have used a single-zeta basis.

\begin{figure*}
\begin{center}
\includegraphics[width=1.5 \columnwidth,angle=-90]{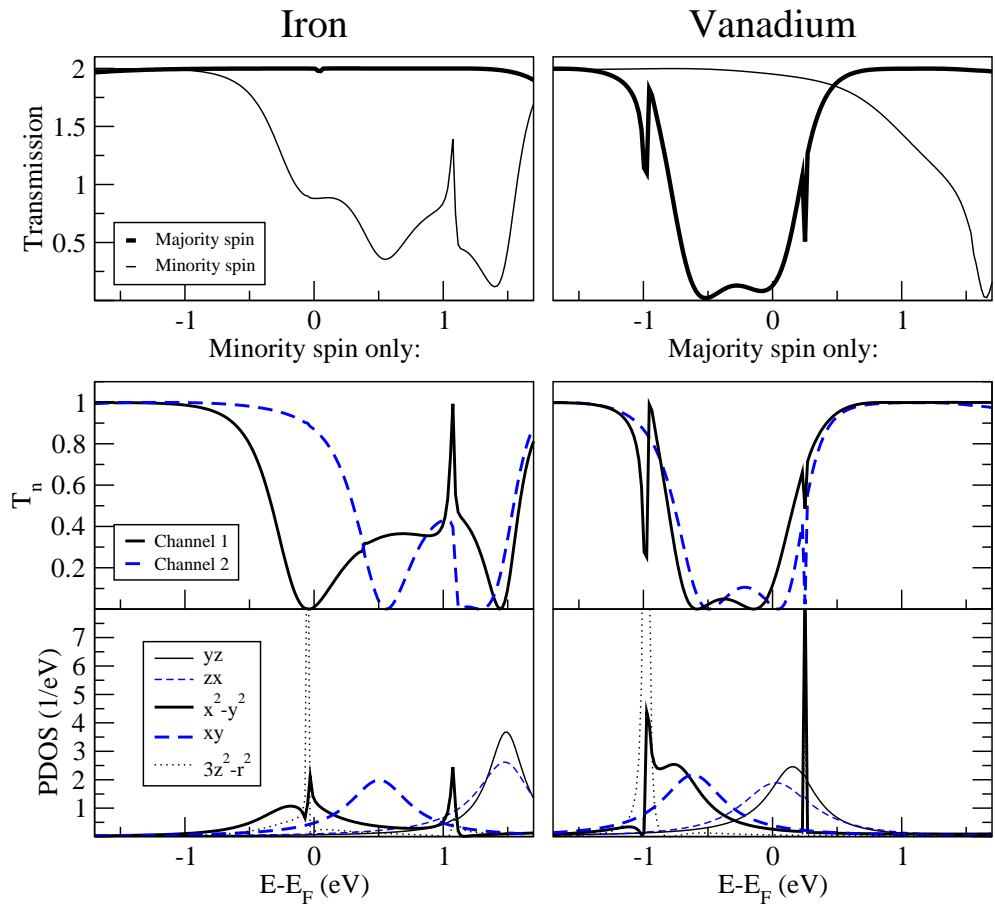}
\caption{Top: Transmission as a function of energy for iron (left)
or vanadium(right). Within a 0.5 eV range of $E_F$ the minority spin channels are suppressed in the
case of iron whereas the majority channels are suppressed for vanadium. Middle: The
transmission of the two minority (majority) spin channels for the iron (vanadium) system. The sum of the two channels yield to total transmission in each case. 
Bottom: The projected density of states (PDOS) of the $3d$ orbitals of the
iron(vanadium) adatom for minority(majority) spin.}\label{slimtrans}
\end{center}
\end{figure*}

\section{Results}\label{sec_three}
The spin resolved transmissions as a function of energy
relative to the Fermi level, $E_F$, of the graphene
sheet with iron and vanadium adatoms are shown in Fig. \ref{slimtrans}.
The transmission of a pure sheet is 2 for each spin type,
since there are two bands in the energy window each contributing with a fully transmitting channel for each spin.
It is seen that spin dependent scattering occurs due to the presence of the adatoms.
In the case of iron the minority spin type is significantly suppressed around $E_{F}$,
whereas the majority spin electrons transmit completely.
For vanadium, likewise, scattering occurs for only one spin channel, but in this case it is the majority spin electrons which are scattered. The transmission of the two bands for minority (iron) and majority (vanadium) spin is shown on Fig. \ref{slimtrans}, middle graph. We see that each band closes completely at certain energies.

In the case of vanadium ($3d^3 4s^2$) we expect from Hund's rules a total spin of 3 and
the majority $d$-states will be located around $E_F$, whereas the
minority states are all empty and well above $E_F$. The Mulliken analysis indicates a half filling of the majority spin $3d_{yz}$ and $3d_{zx}$ orbitals and a full filling of $3d_{x^2-y^2}$, $3d_{xy}$ and $3d_{3z^2-r^2}$. For minority electrons all $3d$ orbitals are empty. This is in full correspondence with the $3d$ orbital PDOS plot shown in Fig. \ref{slimtrans} bottom graph. In the case of iron ($3d^6 4s^2$) we expect from Hund's rules that the majority states are all filled and well below $E_F$, and now the partially filled minority states are located around $E_F$ yielding a total spin of 4. This is again supported by the Mulliken analysis data as well as the projected density of states (PDOS).
\begin{figure*}[t]
\begin{center}
\includegraphics[scale=0.9]{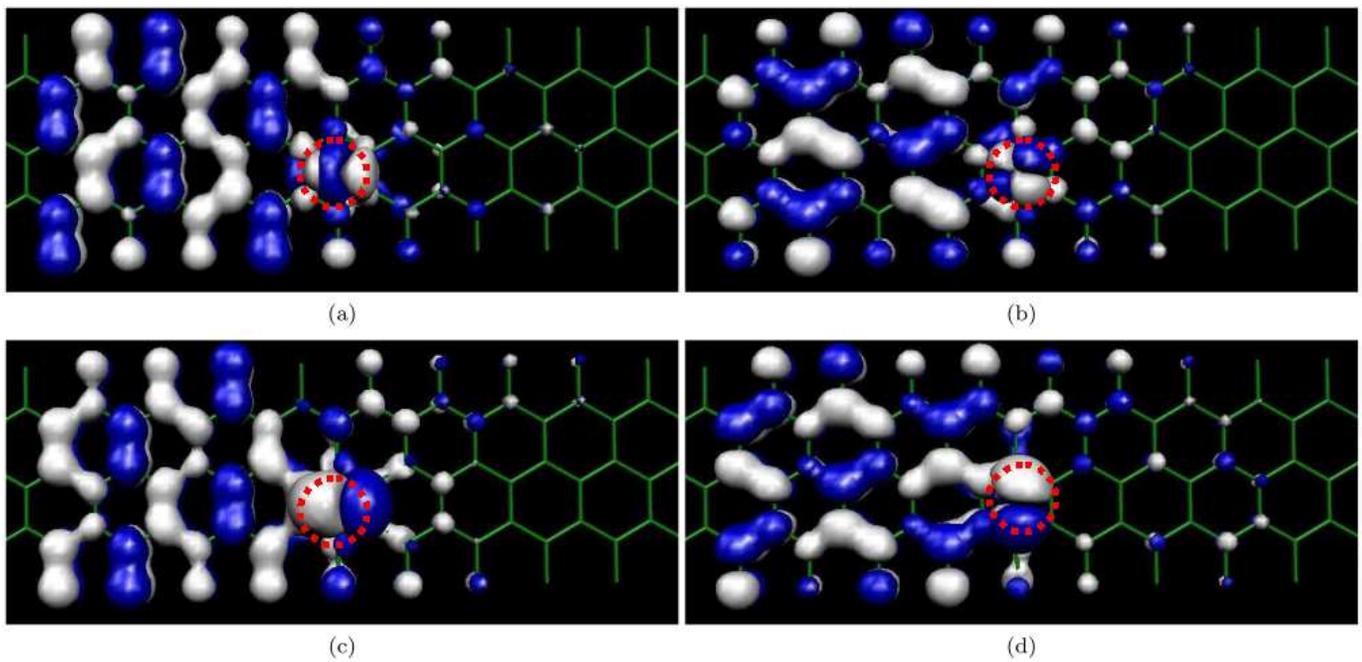}
\caption{(Colour online) The real part of the
(left-to-right) eigenchannel scattering states for the vanadium system, stemming from majority spin electrons
from both bands, at the energy of the first dip (-0.5 eV) (a),(b) and second dip (-0.1 eV) (c),(d) in the
transmission spectrum. Only the solutions in the scattering region are shown.
The size of the shapes indicates a cut-off value for the wavefunction.
White indicates a positive sign a blue a negative sign of the wavefunction.
The involved orbitals of the vanadium atom (marked with dotted circle) are seen to be (a) : $3d_{x^2-y^2}$ (and a minor presence of $3d_{3z^2-r^2}$), (b) :
$3d_{xy}$, (c): $3d_{yz}$ and (d) : $3d_{zx}$.}\label{eigchV}
\end{center}
\end{figure*}
Comparing PDOS and the transmission there is a strong correlation between adatom orbital energies and conductance dips. This is very clear for the two separated channel closings for vanadium at $E \approx E_F$ and $E-E_F\approx -0.5$ eV.
The interference between a waves involving the quasi-bound $d$-state on the adatom and directly transmitted band states yield a Fano (anti-) resonance. The line shape of the resonance is given by the Fano function \cite{FANO1961}
\beq
f(\epsilon)=\frac{(\epsilon + q)^{2}}{\epsilon^{2} + 1}, \label{fanoeq}
\eeq
where $\epsilon=(E-E_{R})/\Gamma$. Here $\Gamma$ is the resonance width, $E_{R}$ the resonance energy and \textit{q} the asymmetry parameter. Despite the overlapping of dips in our calculations their shape appears symmetric in the vicinity of zero transmission. Since the structure has inversion symmetry around the adatom in the transport direction the parameter $q$ must be real and thus we expect to have zero transmission dips as seen in Fig. \ref{slimtrans} at the anti-resonances $\epsilon=-q$\cite{NockelStone,Rau2004}. 
A very sharp resonance can be seen for vanadium around $-1$eV which should also go to zero. However, it is related to the most localized $d$-orbital, $d_{3d^2-r^2}$, and a small asymmetry in the numerical calculation and the resolution cause it not to go exactly to zero. Signatures of Fano resonances have previously been observed experimentally in the conductance of multi-wall carbon nanotubes at low temperature\cite{zhang2}.
  
The anti-resonances are illustrated further by performing an eigenchannel analysis where scattering states with well-defined transmissions ("eigenchannels") are constructed\cite{Paulsson2007}. We take vanadium as our example.
The eigenchannel transmissions, $T_n(\epsilon)$, provide the transmission of channel $n$ at
a given energy $\epsilon$. Plotted on Fig. \ref{eigchV} (a)(b) and (c)(d) are the real part of
the majority spin scattering state solutions of both bands for energies corresponding to the
first dip ($-0.5$eV) and the second dip ($-0.1$eV), respectively. 
We see that the wavefunction is nonvanishing only on the left side of the vanadium atom as expected
since we have full reflection in all cases. The anti-symmetric and symmetric solutions are seen on
Fig. \ref{eigchV} (a)(d) and (b)(c), respectively. Note that in the transverse direction the neighbouring C atoms have the same (opposite)
sign in the symmetric (anti-symmetric) case. These solutions are indeed matched by the vanadium atom
orbitals, which is seen on the figures by a match of signs (colours).
From the shape and sign in the plots the orbitals involved can be identified for each band at both energies.\\

\section{A tight-binding model}\label{sec_four}
\begin{figure}[tbh]
\begin{center}
\includegraphics[width=0.9 \columnwidth,angle=0]{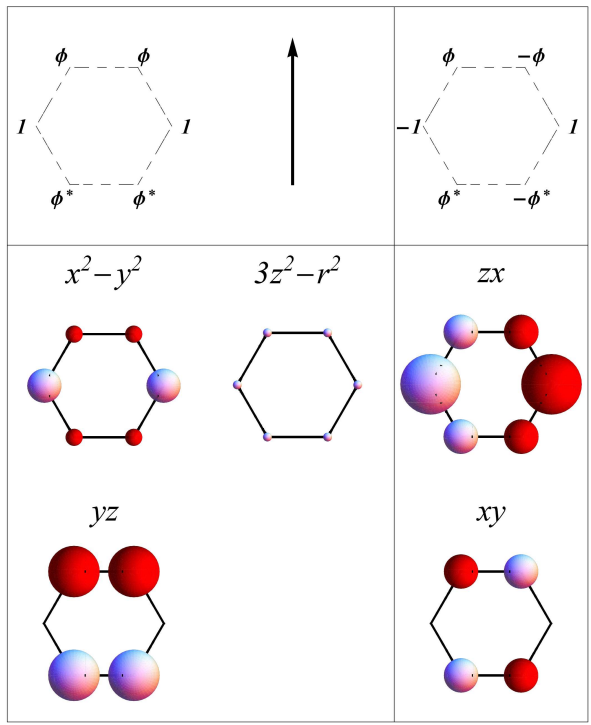}
\end{center}
\caption{(Colour online) Top panel: The two $\pi$-band ($p_z$) states (not normalized) in a
hexagon where the arrow denotes the direction along the armchair tube ($y$
direction) with $\phi=e^{i2\pi/3}$. Left/right panel
correspond to symmetric/anti-symmetric solution on the A,B-dimers ($x$ direction). Lower panels: Simple tight-binding model. The sign and relative size
of the hopping matrix elements between the ring $\pi$-orbitals and the metal adatom
$d$-orbitals when the metal-atom is situated 1.7{\AA~} above the middle of the ring.}
\label{fig.simple1}
\end{figure}

We will now rationalize the results for the channels and transmissions in terms of
the simplest possible tight-binding model. We consider only the coupling of the adatom
$d$-orbitals with the 6 nearest $\pi$-orbitals. The starting point will be the two band states
of the $\pi$-electrons in the armchair direction at the Fermi level, see
Fig.~\ref{fig.simple1}. These are characterized by rotational symmetry around the tube
axis and come in an odd/even version around the symmetry plane normal to $x$ along the
tube ($y$) and cutting through the adatom. They couple to different $d$-orbitals
on the adatom with the same symmetry. Thus the symmetric/anti-symmetric band only
couple to the $d$-orbitals even/odd in $x$.
To make a minimal model (with a minimum of parameters) we assume that the $d$-orbitals have
the same on-site energy $E_d$, and a coupling to the carbon $\pi$-orbitals described in
units of $V_{pd}=V_{pd\sigma}$ taking $V_{pd\pi}\approx 0.5V_{pd\sigma}$
\cite{Harrison1989}. The size and sign of the coupling of the different $d$-orbitals to
the $p_z$ orbitals are illustrated in Fig.~\ref{fig.simple1}, left/right panels
corresponding to coupling to the two types of bands. For both bands it is seen that the
$d_{3z^2-r^2}$ orbitals do not couple at $E_F$ since the wavefunction values around the
hexagon sums up to zero. Therefore we do not expect
significant contributions from this orbital, which will be very localised. 

\begin{figure}[tbh]
\begin{center}
\includegraphics[width=0.9 \columnwidth,angle=-90,viewport=0 0 550 575,clip]{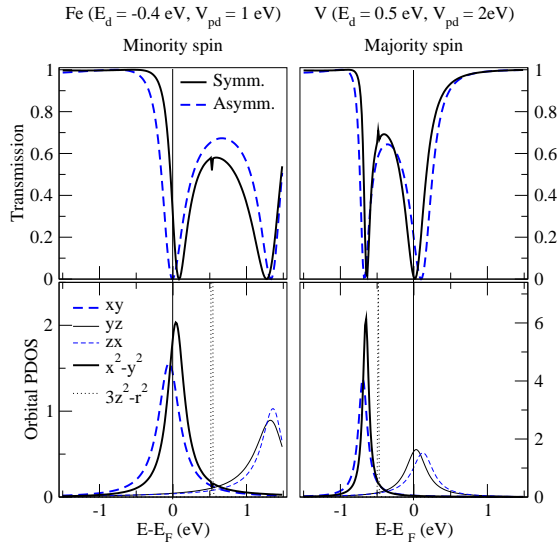}
\end{center}
\caption{ (Colour online) The transmission through the symmetric/anti-symmetric bands for Fe
minority spin (left panels) and V majority spin (right panels) calculated with the simple tight-binding model. Lower panels display the
projected density of states onto the $d$-orbitals.}  \label{fig.simple2}
\end{figure}

We only consider the majority spin $d$-states for V, since they, as mentioned above, will be located around $E_F$ with $E_d\approx 0$. Likewise, for iron we only consider the minority spin states. The model calculation with ($E_d,V_{pd}$) parameters chosen to
fit the data from the full \textit{ab initio} calculation are shown in Fig.~\ref{fig.simple2}. It is seen
that the transmission is more suppressed in the full calculation where the individual
$d$-orbitals are allowed to have different energies, but the fact that we see 4 dips
corresponding to the four coupling $d$-orbitals is clear.

\begin{figure}[tbh]
\begin{center}
\includegraphics[width=0.9 \columnwidth,angle=0,scale=1,viewport=0 0 400 400,clip]{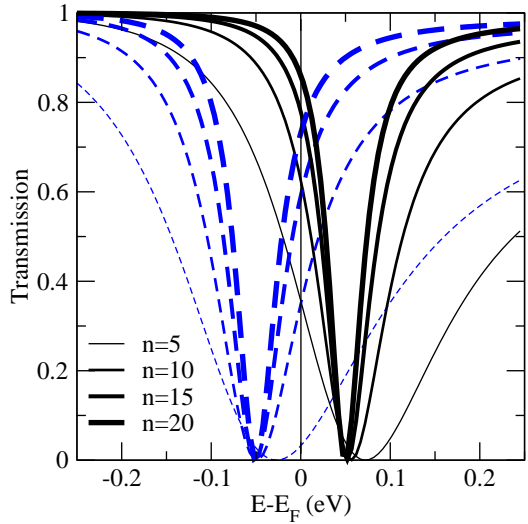}
\end{center}
\caption{(Colour online) The dependence on tube-size, $(n,n)$ calculated with the
simple tight-binding model at the first dip in the Fe case.} \label{fig.simple3}
\end{figure}

With the simple model we can now explore what happens when the
diameter of $(n,n)$-tube is increased. This is shown in Fig. \ref{fig.simple3}
where it is seen that the transmission still goes to zero at the anti-resonance points but in an interval getting narrower with increasing $n$. The latter is in agreement with the findings of Todorov and
White \cite{todorov}. They conclude that the larger the tube diameter the more the
scattering due to an impurity is reduced. This happens because the overlap with the wavefunction (normalised around the tube circumference) with the scattering potential goes down. The fact that the conductance still drops to zero in our calculations is due to the Fano resonance phenomena. 
\section{Conclusion}
We have described spin-polarised zero bias transport calculations for armchair carbon nanotubes with adsorbed
single iron or vanadium atoms. We find a significant difference between transmissions of majority and minority spin. The presence of the metal adatoms
causes spin-dependent closing of the conduction channels at certain energies. The mechanism is due to Fano resonances related to the particular $d$-states close to the Fermi energy. 
Only $d$-orbitals with a symmetry matching the symmetry of the Bloch band solutions take part in the scattering.
The scattered spin type (minority or majority spin) can be explained by the filling of the $3d$ orbitals via Hund's rule. Results of a simple tight-binding model show that increasing the width of the tube, and thus reducing the concentration of adatoms, results in narrower dips in the
conduction, but still with complete closing. The latter is explained by the reflection symmetry of the system in the transport direction.
\begin{acknowledgments}
The authors would like to thank Jeremy Taylor for useful discussions. Computational resources were provided by the Danish Center for Scientific Computations (DCSC). APJ is grateful to the FiDiPro program of
the Finnish Academy for support during the final stages
of this work.
\end{acknowledgments}


\begin{thebibliography}{33}
\expandafter\ifx\csname natexlab\endcsname\relax\def\natexlab#1{#1}\fi
\expandafter\ifx\csname bibnamefont\endcsname\relax
  \def\bibnamefont#1{#1}\fi
\expandafter\ifx\csname bibfnamefont\endcsname\relax
  \def\bibfnamefont#1{#1}\fi
\expandafter\ifx\csname citenamefont\endcsname\relax
  \def\citenamefont#1{#1}\fi
\expandafter\ifx\csname url\endcsname\relax
  \def\url#1{\texttt{#1}}\fi
\expandafter\ifx\csname urlprefix\endcsname\relax\def\urlprefix{URL }\fi
\providecommand{\bibinfo}[2]{#2}
\providecommand{\eprint}[2][]{\url{#2}}

\bibitem[{\citenamefont{Avouris et~al.}(2007)\citenamefont{Avouris, Chen, and
  Perebeinos}}]{Avouris2007}
\bibinfo{author}{\bibfnamefont{P.}~\bibnamefont{Avouris}},
  \bibinfo{author}{\bibfnamefont{Z.}~\bibnamefont{Chen}}, \bibnamefont{and}
  \bibinfo{author}{\bibfnamefont{V.}~\bibnamefont{Perebeinos}},
  \bibinfo{journal}{Nature Nanotechnology} \textbf{\bibinfo{volume}{2}},
  \bibinfo{pages}{605} (\bibinfo{year}{2007}).

\bibitem[{\citenamefont{Geim and Novoselov}(2007)}]{transistor}
\bibinfo{author}{\bibfnamefont{A.}~\bibnamefont{Geim}} \bibnamefont{and}
  \bibinfo{author}{\bibfnamefont{K.}~\bibnamefont{Novoselov}},
  \bibinfo{journal}{Nature Materials} \textbf{\bibinfo{volume}{6}},
  \bibinfo{pages}{183} (\bibinfo{year}{2007}).

\bibitem[{\citenamefont{Prinz}(1998)}]{prinz}
\bibinfo{author}{\bibfnamefont{G.}~\bibnamefont{Prinz}},
  \bibinfo{journal}{Science} \textbf{\bibinfo{volume}{282}},
  \bibinfo{pages}{1660} (\bibinfo{year}{1998}).

\bibitem[{\citenamefont{Wolf}(2001)}]{wolf}
\bibinfo{author}{\bibfnamefont{S.}~\bibnamefont{Wolf}},
  \bibinfo{journal}{Science} \textbf{\bibinfo{volume}{292}},
  \bibinfo{pages}{1488} (\bibinfo{year}{2001}).

\bibitem[{\citenamefont{Zutic et~al.}(2004)\citenamefont{Zutic, Fabian, and
  Das~Sarma}}]{Zutic}
\bibinfo{author}{\bibfnamefont{I.}~\bibnamefont{Zutic}},
  \bibinfo{author}{\bibfnamefont{J.}~\bibnamefont{Fabian}}, \bibnamefont{and}
  \bibinfo{author}{\bibfnamefont{S.}~\bibnamefont{Das~Sarma}},
  \bibinfo{journal}{Rev. Mod. Phys.} \textbf{\bibinfo{volume}{76}},
  \bibinfo{pages}{323} (\bibinfo{year}{2004}).

\bibitem[{\citenamefont{Lehtinen et~al.}(2004)\citenamefont{Lehtinen, Foster,
  Ma, Krasheninnikov, and Nieminen}}]{Lehtinen2004}
\bibinfo{author}{\bibfnamefont{P.}~\bibnamefont{Lehtinen}},
  \bibinfo{author}{\bibfnamefont{A.}~\bibnamefont{Foster}},
  \bibinfo{author}{\bibfnamefont{Y.}~\bibnamefont{Ma}},
  \bibinfo{author}{\bibfnamefont{A.}~\bibnamefont{Krasheninnikov}},
  \bibnamefont{and} \bibinfo{author}{\bibfnamefont{R.}~\bibnamefont{Nieminen}},
  \bibinfo{journal}{Phys. Rev. Lett.} \textbf{\bibinfo{volume}{93}},
  \bibinfo{pages}{187202} (\bibinfo{year}{2004}).

\bibitem[{\citenamefont{Kumazaki and Hirashima}(2007)}]{kumazaki}
\bibinfo{author}{\bibfnamefont{H.}~\bibnamefont{Kumazaki}} \bibnamefont{and}
  \bibinfo{author}{\bibfnamefont{D.~S.} \bibnamefont{Hirashima}},
  \bibinfo{journal}{Journal of the Physical Society of Japan}
  \textbf{\bibinfo{volume}{76}}, \bibinfo{pages}{64713} (\bibinfo{year}{2007}).

\bibitem[{\citenamefont{Duplock et~al.}(2004)\citenamefont{Duplock, Scheffler,
  and Lindan}}]{Duplock2004}
\bibinfo{author}{\bibfnamefont{E.~J.} \bibnamefont{Duplock}},
  \bibinfo{author}{\bibfnamefont{M.}~\bibnamefont{Scheffler}},
  \bibnamefont{and} \bibinfo{author}{\bibfnamefont{P.~J.}
  \bibnamefont{Lindan}}, \bibinfo{journal}{Phys. Rev. Lett.}
  \textbf{\bibinfo{volume}{92}}, \bibinfo{pages}{225502}
  (\bibinfo{year}{2004}).

\bibitem[{\citenamefont{Okada and Oshiyama}(2001)}]{okada}
\bibinfo{author}{\bibfnamefont{S.}~\bibnamefont{Okada}} \bibnamefont{and}
  \bibinfo{author}{\bibfnamefont{A.}~\bibnamefont{Oshiyama}},
  \bibinfo{journal}{Phys. Rev. Lett.} \textbf{\bibinfo{volume}{87}},
  \bibinfo{pages}{146803} (\bibinfo{year}{2001}).

\bibitem[{\citenamefont{Lee et~al.}(2005)\citenamefont{Lee, Son, Park, Han, and
  Yu}}]{lee}
\bibinfo{author}{\bibfnamefont{H.}~\bibnamefont{Lee}},
  \bibinfo{author}{\bibfnamefont{Y.-W.} \bibnamefont{Son}},
  \bibinfo{author}{\bibfnamefont{N.}~\bibnamefont{Park}},
  \bibinfo{author}{\bibfnamefont{S.}~\bibnamefont{Han}}, \bibnamefont{and}
  \bibinfo{author}{\bibfnamefont{J.}~\bibnamefont{Yu}}, \bibinfo{journal}{Phys.
  Rev. B} \textbf{\bibinfo{volume}{72}}, \bibinfo{pages}{174431}
  (\bibinfo{year}{2005}).

\bibitem[{\citenamefont{Fagan et~al.}(2003{\natexlab{a}})\citenamefont{Fagan,
  Mota, da~Silva, and Fazzio}}]{fagan3}
\bibinfo{author}{\bibfnamefont{B.}~\bibnamefont{Fagan}},
  \bibinfo{author}{\bibfnamefont{R.}~\bibnamefont{Mota}},
  \bibinfo{author}{\bibfnamefont{A.}~\bibnamefont{da~Silva}}, \bibnamefont{and}
  \bibinfo{author}{\bibfnamefont{A.}~\bibnamefont{Fazzio}},
  \bibinfo{journal}{Physica B} \textbf{\bibinfo{volume}{340}},
  \bibinfo{pages}{982} (\bibinfo{year}{2003}{\natexlab{a}}).

\bibitem[{\citenamefont{Fagan et~al.}(2004)\citenamefont{Fagan, Mota, da~Silva,
  and Fazzio}}]{fagan2}
\bibinfo{author}{\bibfnamefont{B.}~\bibnamefont{Fagan}},
  \bibinfo{author}{\bibfnamefont{R.}~\bibnamefont{Mota}},
  \bibinfo{author}{\bibfnamefont{A.}~\bibnamefont{da~Silva}}, \bibnamefont{and}
  \bibinfo{author}{\bibfnamefont{A.}~\bibnamefont{Fazzio}},
  \bibinfo{journal}{J.Phys.:Condens. Matter} \textbf{\bibinfo{volume}{16}},
  \bibinfo{pages}{3647} (\bibinfo{year}{2004}).

\bibitem[{\citenamefont{Fagan et~al.}(2003{\natexlab{b}})\citenamefont{Fagan,
  Mota, da~Silva, and Fazzio}}]{fagan1}
\bibinfo{author}{\bibfnamefont{S.~B.} \bibnamefont{Fagan}},
  \bibinfo{author}{\bibfnamefont{R.}~\bibnamefont{Mota}},
  \bibinfo{author}{\bibfnamefont{A.~J.~R.} \bibnamefont{da~Silva}},
  \bibnamefont{and} \bibinfo{author}{\bibfnamefont{A.}~\bibnamefont{Fazzio}},
  \bibinfo{journal}{Phys. Rev. B} \textbf{\bibinfo{volume}{67}},
  \bibinfo{pages}{205414} (\bibinfo{year}{2003}{\natexlab{b}}).

\bibitem[{\citenamefont{Duffy and Blackman}(1998)}]{Duffy1998}
\bibinfo{author}{\bibfnamefont{D.}~\bibnamefont{Duffy}} \bibnamefont{and}
  \bibinfo{author}{\bibfnamefont{J.}~\bibnamefont{Blackman}},
  \bibinfo{journal}{Phys. Rev. B} \textbf{\bibinfo{volume}{58}},
  \bibinfo{pages}{7443} (\bibinfo{year}{1998}).

\bibitem[{\citenamefont{Yang et~al.}(2003)\citenamefont{Yang, Zhao, and
  Lu}}]{Yang}
\bibinfo{author}{\bibfnamefont{C.-K.} \bibnamefont{Yang}},
  \bibinfo{author}{\bibfnamefont{J.}~\bibnamefont{Zhao}}, \bibnamefont{and}
  \bibinfo{author}{\bibfnamefont{J.~P.} \bibnamefont{Lu}},
  \bibinfo{journal}{Phys. Rev. Lett.} \textbf{\bibinfo{volume}{90}},
  \bibinfo{pages}{257203} (\bibinfo{year}{2003}).

\bibitem[{\citenamefont{Kang et~al.}(2005)\citenamefont{Kang, Choi, Moon, and
  Chang}}]{6}
\bibinfo{author}{\bibfnamefont{Y.-J.} \bibnamefont{Kang}},
  \bibinfo{author}{\bibfnamefont{J.}~\bibnamefont{Choi}},
  \bibinfo{author}{\bibfnamefont{C.-Y.} \bibnamefont{Moon}}, \bibnamefont{and}
  \bibinfo{author}{\bibfnamefont{K.}~\bibnamefont{Chang}},
  \bibinfo{journal}{Phys. Rev. B} \textbf{\bibinfo{volume}{71}},
  \bibinfo{pages}{115441} (\bibinfo{year}{2005}).

\bibitem[{\citenamefont{Kishi et~al.}(2007)\citenamefont{Kishi, David, Wilson,
  Nakanishi, and Kasai}}]{kishi}
\bibinfo{author}{\bibfnamefont{T.}~\bibnamefont{Kishi}},
  \bibinfo{author}{\bibfnamefont{M.}~\bibnamefont{David}},
  \bibinfo{author}{\bibfnamefont{A.}~\bibnamefont{Wilson}},
  \bibinfo{author}{\bibfnamefont{H.}~\bibnamefont{Nakanishi}},
  \bibnamefont{and} \bibinfo{author}{\bibfnamefont{H.}~\bibnamefont{Kasai}},
  \bibinfo{journal}{The Jap. Journal of Appl. Phys.}
  \textbf{\bibinfo{volume}{46}}, \bibinfo{pages}{1788} (\bibinfo{year}{2007}).

\bibitem[{\citenamefont{Monthioux}(2002)}]{monthioux}
\bibinfo{author}{\bibfnamefont{M.}~\bibnamefont{Monthioux}},
  \bibinfo{journal}{Carbon} \textbf{\bibinfo{volume}{40}},
  \bibinfo{pages}{1809} (\bibinfo{year}{2002}).

\bibitem[{\citenamefont{Zhang et~al.}(2006)\citenamefont{Zhang, Wang, and
  Cheng}}]{zhang}
\bibinfo{author}{\bibfnamefont{C.}~\bibnamefont{Zhang}},
  \bibinfo{author}{\bibfnamefont{L.}~\bibnamefont{Wang}}, \bibnamefont{and}
  \bibinfo{author}{\bibfnamefont{H.}~\bibnamefont{Cheng}},
  \bibinfo{journal}{The Journal of Chem. Phys.} \textbf{\bibinfo{volume}{124}},
  \bibinfo{pages}{201107} (\bibinfo{year}{2006}).

\bibitem[{\citenamefont{Wang et~al.}(2007)\citenamefont{Wang, Zhu, Ren, Wang,
  and Guo}}]{guo}
\bibinfo{author}{\bibfnamefont{B.}~\bibnamefont{Wang}},
  \bibinfo{author}{\bibfnamefont{Y.}~\bibnamefont{Zhu}},
  \bibinfo{author}{\bibfnamefont{W.}~\bibnamefont{Ren}},
  \bibinfo{author}{\bibfnamefont{J.}~\bibnamefont{Wang}}, \bibnamefont{and}
  \bibinfo{author}{\bibfnamefont{H.}~\bibnamefont{Guo}},
  \bibinfo{journal}{Phys. Rev. B} \textbf{\bibinfo{volume}{75}},
  \bibinfo{pages}{235415} (\bibinfo{year}{2007}).

\bibitem[{\citenamefont{Martins et~al.}(2007)\citenamefont{Martins, Miwa,
  da~Silva, and Fazzio}}]{martins_dasilva}
\bibinfo{author}{\bibfnamefont{T.}~\bibnamefont{Martins}},
  \bibinfo{author}{\bibfnamefont{R.}~\bibnamefont{Miwa}},
  \bibinfo{author}{\bibfnamefont{A.}~\bibnamefont{da~Silva}}, \bibnamefont{and}
  \bibinfo{author}{\bibfnamefont{A.}~\bibnamefont{Fazzio}},
  \bibinfo{journal}{Phys. Rev. B} \textbf{\bibinfo{volume}{98}},
  \bibinfo{pages}{196803} (\bibinfo{year}{2007}).

\bibitem[{\citenamefont{Soler et~al.}(2002)\citenamefont{Soler, Artacho, Gale,
  Garcia, Junquera, Ordejon, and Sanchez-Portal}}]{siesta}
\bibinfo{author}{\bibfnamefont{J.~M.} \bibnamefont{Soler}},
  \bibinfo{author}{\bibfnamefont{E.}~\bibnamefont{Artacho}},
  \bibinfo{author}{\bibfnamefont{J.~D.} \bibnamefont{Gale}},
  \bibinfo{author}{\bibfnamefont{A.}~\bibnamefont{Garcia}},
  \bibinfo{author}{\bibfnamefont{J.}~\bibnamefont{Junquera}},
  \bibinfo{author}{\bibfnamefont{P.}~\bibnamefont{Ordejon}}, \bibnamefont{and}
  \bibinfo{author}{\bibfnamefont{D.}~\bibnamefont{Sanchez-Portal}},
  \bibinfo{journal}{Journal Of Physics-Condensed Matter}
  \textbf{\bibinfo{volume}{14}}, \bibinfo{pages}{2745} (\bibinfo{year}{2002}).

\bibitem[{\citenamefont{Perdew et~al.}(1996)\citenamefont{Perdew, Burke, and
  Ernzerhof}}]{Perdew1996}
\bibinfo{author}{\bibfnamefont{J.~P.} \bibnamefont{Perdew}},
  \bibinfo{author}{\bibfnamefont{K.}~\bibnamefont{Burke}}, \bibnamefont{and}
  \bibinfo{author}{\bibfnamefont{M.}~\bibnamefont{Ernzerhof}},
  \bibinfo{journal}{Phys. Rev. Lett.} \textbf{\bibinfo{volume}{77}},
  \bibinfo{pages}{3865} (\bibinfo{year}{1996}).

\bibitem[{\citenamefont{Brandbyge et~al.}(2002)\citenamefont{Brandbyge, Mozos,
  Ordejon, Taylor, and Stokbro}}]{Brandbyge2002}
\bibinfo{author}{\bibfnamefont{M.}~\bibnamefont{Brandbyge}},
  \bibinfo{author}{\bibfnamefont{J.-L.} \bibnamefont{Mozos}},
  \bibinfo{author}{\bibfnamefont{P.}~\bibnamefont{Ordejon}},
  \bibinfo{author}{\bibfnamefont{J.}~\bibnamefont{Taylor}}, \bibnamefont{and}
  \bibinfo{author}{\bibfnamefont{K.}~\bibnamefont{Stokbro}},
  \bibinfo{journal}{Phys. Rev. B} \textbf{\bibinfo{volume}{65}},
  \bibinfo{pages}{165401/1} (\bibinfo{year}{2002}).

\bibitem[{\citenamefont{Datta}(1997)}]{Datta1997}
\bibinfo{author}{\bibfnamefont{S.}~\bibnamefont{Datta}},
  \emph{\bibinfo{title}{Electronic Transport in Mesoscopic Systems}}
  (\bibinfo{publisher}{Cambridge University Press, Cambridge, England},
  \bibinfo{year}{1997}).

\bibitem[{\citenamefont{White and Mintmire}(2005)}]{White2005}
\bibinfo{author}{\bibfnamefont{C.~T.} \bibnamefont{White}} \bibnamefont{and}
  \bibinfo{author}{\bibfnamefont{J.~W.} \bibnamefont{Mintmire}},
  \bibinfo{journal}{Journal of Physical Chemistry B}
  \textbf{\bibinfo{volume}{109}}, \bibinfo{pages}{52} (\bibinfo{year}{2005}).

\bibitem[{\citenamefont{Fano}(1961)}]{FANO1961}
\bibinfo{author}{\bibfnamefont{U.}~\bibnamefont{Fano}}, \bibinfo{journal}{Phys.
  Rev.} \textbf{\bibinfo{volume}{1}}, \bibinfo{pages}{1866}
  (\bibinfo{year}{1961}).

\bibitem[{\citenamefont{Nockel and Stone}(1994)}]{NockelStone}
\bibinfo{author}{\bibfnamefont{J.~U.} \bibnamefont{Nockel}} \bibnamefont{and}
  \bibinfo{author}{\bibfnamefont{A.~D.} \bibnamefont{Stone}},
  \bibinfo{journal}{Phys. Rev. B} \textbf{\bibinfo{volume}{50}},
  \bibinfo{pages}{17415} (\bibinfo{year}{1994}).

\bibitem[{\citenamefont{Rau}(2004)}]{Rau2004}
\bibinfo{author}{\bibfnamefont{A.~R.~P.} \bibnamefont{Rau}},
  \bibinfo{journal}{Physica Scripta} \textbf{\bibinfo{volume}{69}},
  \bibinfo{pages}{C10} (\bibinfo{year}{2004}).

\bibitem[{\citenamefont{Zhang and Chandrasekhar}(2006)}]{zhang2}
\bibinfo{author}{\bibfnamefont{Z.}~\bibnamefont{Zhang}} \bibnamefont{and}
  \bibinfo{author}{\bibfnamefont{V.}~\bibnamefont{Chandrasekhar}},
  \bibinfo{journal}{Phys. Rev. B} \textbf{\bibinfo{volume}{73}},
  \bibinfo{eid}{075421} (pages~\bibinfo{numpages}{10}) (\bibinfo{year}{2006}).

\bibitem[{\citenamefont{Paulsson and Brandbyge}(2007)}]{Paulsson2007}
\bibinfo{author}{\bibfnamefont{M.}~\bibnamefont{Paulsson}} \bibnamefont{and}
  \bibinfo{author}{\bibfnamefont{M.}~\bibnamefont{Brandbyge}},
  \bibinfo{journal}{Phys. Rev. B} \textbf{\bibinfo{volume}{76}},
  \bibinfo{pages}{115117} (\bibinfo{year}{2007}).

\bibitem[{\citenamefont{Harrison}(1989)}]{Harrison1989}
\bibinfo{author}{\bibfnamefont{W.~A.} \bibnamefont{Harrison}},
  \emph{\bibinfo{title}{Electronic Structure and the Properties of Solids}}
  (\bibinfo{publisher}{Dover}, \bibinfo{year}{1989}).

\bibitem[{\citenamefont{White and Todorov}(1998)}]{todorov}
\bibinfo{author}{\bibfnamefont{C.~T.} \bibnamefont{White}} \bibnamefont{and}
  \bibinfo{author}{\bibfnamefont{T.~N.} \bibnamefont{Todorov}},
  \bibinfo{journal}{Nature} \textbf{\bibinfo{volume}{393}},
  \bibinfo{pages}{240} (\bibinfo{year}{1998}).

\end{thebibliography}


\begin{thebibliography}{18}
\expandafter\ifx\csname natexlab\endcsname\relax\def\natexlab#1{#1}\fi
\expandafter\ifx\csname bibnamefont\endcsname\relax
  \def\bibnamefont#1{#1}\fi
\expandafter\ifx\csname bibfnamefont\endcsname\relax
  \def\bibfnamefont#1{#1}\fi
\expandafter\ifx\csname citenamefont\endcsname\relax
  \def\citenamefont#1{#1}\fi
\expandafter\ifx\csname url\endcsname\relax
  \def\url#1{\texttt{#1}}\fi
\expandafter\ifx\csname urlprefix\endcsname\relax\def\urlprefix{URL }\fi
\providecommand{\bibinfo}[2]{#2}
\providecommand{\eprint}[2][]{\url{#2}}\bibliographystyle{unsrt}
\bibliography{./hej}

\end{thebibliography}
\end{document}